\documentclass[12pt,3p]{elsarticle}
\usepackage{graphicx,wrapfig,color,amsmath,subfigure}
\usepackage{hyperref}
\title{Noise Characterization in Free Space Polarization Modulation Communication Using Simulated Atmospheric Conditions in Laboratory}
\author{Ram Soorat and Ashok Vudayagiri \\ School of Physics, University of Hyderabad
Hyderabad, 500046 India }
\ead{avsp@uohyd.ernet.in}

\begin{document}

\begin{abstract}
We present a practical scheme for measurement-device-independent polarization shift keying using two state polarization encoding. Most of the previous work on optical free space laser communications through the atmosphere was concentrated on intensity modulated systems. However, polarization modulated systems may be more appropriate for such communication links, because the polarization seems to be the most stable characteristic of a laser beam while propagating through the atmosphere. Thus, a detailed comparison between intensity and polarization modulated systems is of much interest. We analyse the noise in presence of simulated smoke and fog conditions within laboratory and propose a practical scheme for extracting message from the received data. The proposed method uses only two detectors to analyse the polarizations and the practical definition of state of polarization enables a higher signal-to-noise ratio even in presence of depolaration elements in atmosphere such as fog and smoke. The system also takes into account existing imperfections within the experimental setup and hence is more robust. 
\end{abstract}

\maketitle

\section{Introduction}
Modern day communication technology are increasingly adopting optical technologies, including the last mile.  Most of them are fibre based and use ON-Off keying or Phase shift encoding (PSK). But there has also been some interest in free space communication methods, particularly in situations where fibres can not be laid, such as between moving parties or satellite based communication. 

Free Space Optical (FSO) link is in general a line-of-sight (LOS) setup, where the transmitter and the receiver should directly see one another without any obstructions in between \cite{plank,rez,niu}. Apart from intensity and phase modulation, FSO links can also use polarization modulations unlike a fibre based system.  We show in this communication that polarization modulation is relatively more robust than On/Off keying. In addition, phase shift keying require techniques such as homodyne detection, or coherent detection, leading to further technical consideration. Polarization modulation, as presented here requires a more direct measurement, thus simplifying the technology. 
FSO has numerous applications, ranging from short range interconnects such as  on-chip clock and data transmissions, to outdoor intra-building or even intra-satellite and earth-to-satellite links  \cite{plank}-\cite{noor}. 
FSO becomes particularly useful for when one or both parties involved are mobile links. 

However, all terrestrial FSO links suffer from weather related issues, irrespective of its modulation scheme. Attenuation due to fog, smoke and turbulence affects the link and causes errors \cite{dadrasnia,smyth,zabidi,andrews,majumdar}. In addition,  polarization suffers from scrambling effects due to multiple scattering events through fog or smoke. Intuition  therefore suggests that PolSK is not really a reliable method for free space communication through atmospheric scatterers. But we show below by creating atmospheric effects inside the lab, that polarization degradation is not as harmful as expected and information can still be obtained. 

\section{The concept behind the scheme}
\begin{figure}[h]
\includegraphics[scale=0.5]{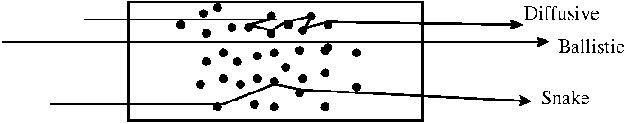}
\caption{Schematic showing ballistic, diffusive and snake photons from a multiple scatterer media.}
\label{snake_photons}
\end{figure}

Atmospheric phenomena such as fog, smoke etc. are essentially suspended particles in air which scatters the light coming from the transmitter. Much of this light is scattered into random directions and do not reach the detector at all, causing a strong attenuation of the signal. Some part does reach the detector, when multiple scattering events eventually lead it towards the detector (see figure \ref{snake_photons}). However, due to random nature of this scattering, this part of light is completely depolarized. They are referred to as `diffused' photons. Some lucky photons escape any scattering at all and directly reach the detector and are termed as `ballistic' photons. Some more photons suffer minimal, grazing scatterings and reach the detector and are called `snake' photons. While diffused photons are completely depolarized, snake photons are minimally depolarized and ballistic ones retain their complete polarization \cite{hema}.

Since the information is embedded in polarization of the photons, it is the snake and ballistic photons which are relevant to us. Diffused photons therefore are a part of noise and if more and more of the incident photons fall into the diffused regime, information gets scrambled and will be completely lost. Hence the ratio of ballistic and snake photons to the diffused is equivalent to signal-to-noise ratio, but this quantity depends upon several parameters such as density of scatters as well as scattering cross section etc. A more useful quantity would then be {\em Degree of Polarization}, which is the ratio of intensity of polarized light to the total intensity, which will be a working definition for signal-to-noise ratio.

\subsection{Degree of Polarization and State of Polarization}
Traditionally, partially polarized light is described by the Stokes vector $\vec{s}=\{s_0,s_1,s_2,s_3\}$ and a {\em Degree of Polarization} defined by \cite{wolf}

$${\cal P}=\frac{(s_1^2+s_2^2+s_3^2)}{s_0^2},$$

The value for DOP ranges between 0 and 1, with 0 defining a completely depolarized light and 1 defining a completely polarized light.  However, a true evaluation of DOP through stokes vector involves six different measurements. But as long as our information is restricted to two photon states, say vertical and horizontal, the measurement reduces to two. Some practical definitions can then be obtained as follows. 

Consider a horizontal polarized beam denoted by $E_x={\cal E}_x \exp(i\omega t- i k.z)$ and a vertical polarized beam $E_y={\cal E}_y \exp(i\omega t- i k.z)$, with standard notations. After depolarization due to atmospheric scattering, their respective amplitudes can be written as a sum of polarized and unpolarized parts. 

\begin{eqnarray}
A_x&=&{\cal E}_x+{\cal E}_{unpol}\cr
A_y&=&{\cal E}_y+{\cal E}_{unpol}
\label{amplitude_total}
\end{eqnarray}

Since polarization is equivalent to information, the unpolarized component represents noise. 

We define a `state-of-polarization' given by 

\begin{equation}
{\cal S}=\frac{|{\cal E}_x|^2-|{\cal E}_y|^2}{|{\cal E}_x|^2+|{\cal E}_y|^2}.
\label{sop1}
\end{equation}

${\cal S}$ ranges from $-1$ to $+1$ for a purely plane polarized light. The range is less than $\pm 1$ for partially polarized light and the definition can not be used for circular and elliptical polarized light. This is a reduced definition from the full Stokes vector formalism and represents $\{s_1\}$ part alone, which is enough for our purpose. 

In the subsequent section, we show that equation \ref{sop1} can be rewritten as 

\begin{equation}
{\cal S}=\frac{|A_x|^2-|A_y|^2}{|A_x|^2+|A_y|^2} = \frac{I_x-I_y}{I_x+I_y},
\label{sop2}
\end{equation}

where $I_{x,y}$ indicate the total intensities measured in $x$ and $y$ (Horizontal and Vertical) polarizations. From equation (\ref{amplitude_total}), it is evident that  this term would also include contribution from the depolarized part. However, in our experimental setup, the depolarized part contributes equally to both horizontal and vertical polarized components and hence cancel out in measurement of ${\cal S}$, leaving only the contribution from polarized part. This is part is discussed in detail in following section. 

It may be noted that the equation (\ref{sop2}) is same as the one obtained from the polarization coherence matrix given by Wolf \cite{wolf}, as
$${\bf J}=\frac{1}{N}\left[\begin{matrix}
\langle E_x E_x \rangle & \langle E_x E_y \rangle \cr
\langle E_y E_x \rangle & \langle E_y E_y \rangle 
\end{matrix}\right]
=\left[\begin{matrix}
\rho_{xx} & \rho_{xy} \cr
\rho_{yx}  & \rho_{yy} 
\end{matrix}\right]
$$

where $N=\langle E_x E_x \rangle+\langle E_x E_x \rangle$, is the normalization factor and ${\cal S}=Tr({\bf J})$.

\section{Experiment}
\begin{figure}[!h]
\includegraphics[width=\textwidth]{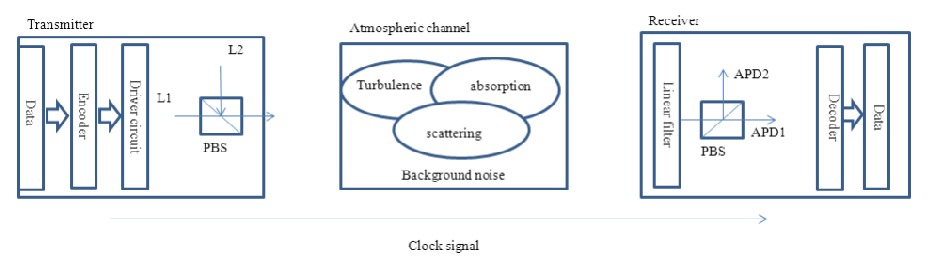}
\caption{Block diagram of experimental setup}
\label{setup}
\end{figure}

Our setup is schematically shown in figure \ref{setup}. The transmitter part consists of two VCSEL's   (VCSEL-780nm from Thorlabs Inc., 1.68 mW power at 780 nm.) labeled L1 and L2. Their output beams are mixed into a single channel using a polarizing beam splitter (PBS), such that vertical component from L1 and horizontal component of L2 are fed into the main communication channel. L1 is pulsed when message bit is 0 and L2 when message bit 1.

The receiver consists of another polarizing beam splitter and two Single Photon Counting Modules (SPCM), based on Avalanche Photodiodes (SensL PCD mini 0200 with sensor area 20 $\mu$m, cooled to -20$^\circ$ C using Peltier modules). They are labelled APD1 and APD2.  The TTL pulses from these units are collected using a DAQ card (NI-PCI-6320, from National Instruments) and saved onto the computer using LabVIEW program. In addition both laser pulses and the counting of APD pulses are synchronized to a clock signal. An interference filter with transmission at 780$\pm$2 nm (FL780-10 from Thorlabs Inc.) ensures that stray light from other sources do not enter the detectors.

Following the equation (\ref{sop2}), the program computes State-of-Polarization (SOP) as

\begin{equation}
{\rm SOP}={\cal S}=\frac{\rm APD1-APD2}{\rm APD1+APD2}
\label{sop_apd}
\end{equation}

where APD1 is a measurement of total vertical polarized component and APD2 is a measurement of total horizontal polarized component. When partially polarized light is incident on the PBS, it can be treated as a sum of polarized and unpolarized components and the polarized part reaches APD1 or APD2 depending upon its polarization. The unpolarized part is split into two components of  equal intensity and orthogonal polarizations and they reach APD1 and APD2 giving rise to equal counts. In other words, output of L1, when reaches PBS would have the intensity $I=(1-p)I_x+pI_{unpol}$, where `p' is the depolarization factor. Out of this intensity, $(1-p)I_x+(p/2)I_{unpol}$ would reach APD1 and $(p/2)I_{unpol}$ would reach APD2. Hence the `state-of-polarization' equation (\ref{sop_apd}) would be,  

\begin{eqnarray}
{\cal S}&=&\frac{\left[(1-p)I_x+(p/2)I_{unpol}\right]-(p/2)I_{unpol}}{(1-p)I_x+pI_{unpol}} \cr \cr
    &=& \frac{(1-p)I_x}{(1-p)I_x+pI_{unpol}}
\end{eqnarray}

Similarly `y' polarized light from L2, incident on PBS  after depolarization would give

\begin{eqnarray}
{\cal S}&=&\frac{(p/2)I_{unpol}-\left[(1-p)I_y+(p/2)I_{unpol}\right]}{(1-p)I_y+pI_{unpol}} \cr \cr
    &=& \frac{-(1-p)I_y}{(1-p)I_y+pI_{unpol}}
\end{eqnarray}

 The initial  `information' at the transmitter is just  vertical or horizontal polarization, and hence message bits 0 and 1 correspond to State of polarization $-1$ and $+1$. At the receiver, this reduces to $-(1-p)$ and $+(1-p)$. The noise factor therefore only reduces the range of SOP rather than completely scrambling it,  as long as the value of $(1-p)$ is small.  The information embedded in horizontal or vertical polarized light then can be obtained, by simply assigning bit 0 whenever SOP is negative and bit 1 whenever SOP is positive.  When $p$ approaches 1, a complete depolarization occurs and the SOP at the receiver is zero. The message in such a case is completely lost. 

\section{Results \& Discusssions}
We present below the results of our experiments. In order to uniformly test the effect of polarization degradation, we initially create a random sequence of 0's and 1's using the LabVIEW's psuedo-random number generator. The sequence is checked for auto-correlation and cross-correlation and both are found to be very near 0.5. These random bits are then mapped to L1 and L2 and transmitted through free space. 

The atmospheric effects are simulated by using a chamber made of glass (35 x 25 x 20 cm) placed in the space between transmitter and the receiver, similar to the setup of Muhammad Ijaz and coworkers \cite{ijaz}. Smoke or fog at various densities are filled into this chamber and the data transmitted through the chamber is analysed. The density of smoke or fog in the path of the communication channel is measured in terms of its effect on the communication, as attenuation of the transmitted light. 

\begin{equation}
{\rm OD (Optical~ Density)} = 10*\log\left(\frac{\rm output~intensity}{\rm incident~intensity}\right)
\label{optical_density}
\end{equation}

 Since attenuation is proportional to density of scattering particles, higher OD also leads to an increased chance of multiple scattering and hence an increase in number of diffused photons and thus a higher polarization noise.  

\begin{figure}[!h]
\subfigure[]{\includegraphics[scale=0.4]{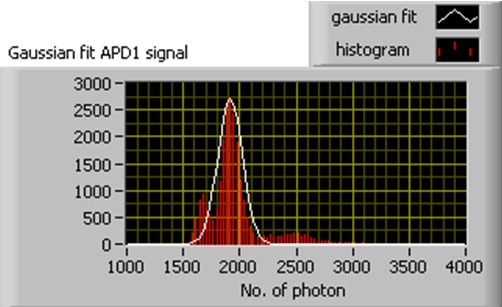}\label{apd11}}\hskip1cm\subfigure[]{\includegraphics[scale=0.4]{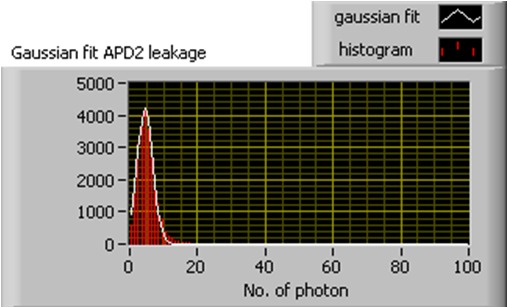}\label{apd12}}\\
\subfigure[]{\includegraphics[scale=0.4]{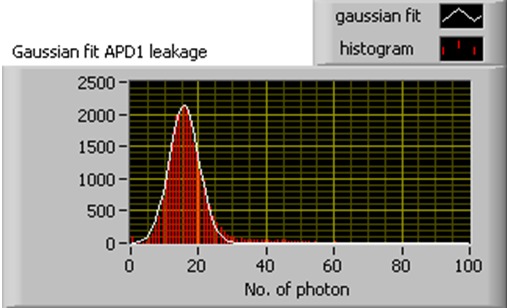}\label{apd21}}\hskip1cm\subfigure[]{\includegraphics[scale=0.4]{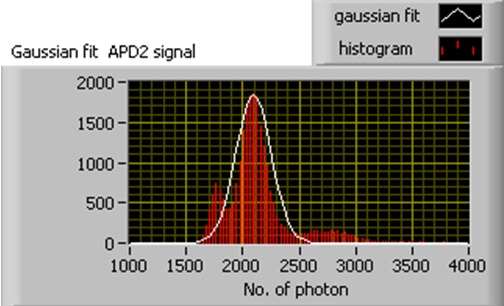}\label{apd22}}
\caption{Histograms for photon counts for calibration data. (a) and (b) are respectively APD1 and APD2 counts when only L1 is used with vertical polarized pulses. While APD1 alone should have registered counts, APD2 also exhibits some counts, though averaged at about 15 as against average of 2000 counts for APD1. Similarly (c) and (d) are for when only L2 is used (horizontal polarization). This background counts (termed `leakage') are due to dark counts, stray light as well as some imperfections in polarizing beam splitters. Since histograms do not overalp, signal can be extracted inspite of these imperfections.}
\label{leakage}
\end{figure}

Prior to calculating the SOP, we calibrated our system by first transmitting only vertical or horizontal polarized pulses and noting down the counts at the detector. When only  horizontal polarized light is incident, only APD2 should show counts. But  we noticed that APD1 also has some counts (figures \ref{apd11} and \ref{apd12}). Similarly, when only vertical polarized pulses are transmitted, APD2 has some background counts even though only APD1 should have shown all the counts. (figures \ref{apd21} and \ref{apd22}). These are due to imperfections in polarizing beamsplitters as well as some dark counts of the APDs, which we term as `leakage'. Despite this, the information is discernible as long as the two histograms of figure \ref{leakage} do not overlap. 

. 

\begin{figure}
\subfigure[]{\includegraphics[scale=0.4]{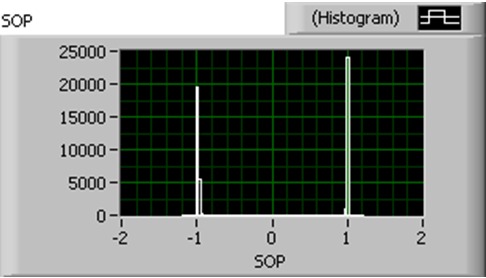}\label{SOP_histogram}}\hskip2cm\subfigure[]{\includegraphics[scale=0.4]{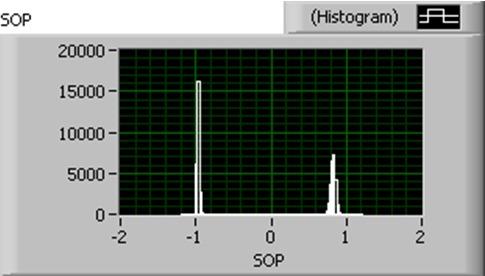}\label{total_sop}}
\caption{(a) Distribution of State of Polarization with no atmospheric effects in between. The distribution consists of two non-overlapping peaks centred at $-1$ and $+1$ respectively. (b) With smoke at OD=-25 dB in path. The range is reduced from $\pm 1$. However, bit assignment will not have errors until the distributions start overlapping.}
\label{SOP_histograms}
\end{figure}

\begin{figure}[h]
\includegraphics[scale=0.4]{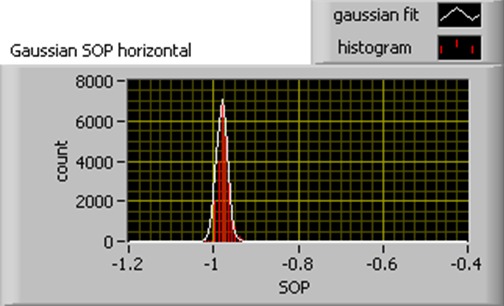}\hskip1.5cm \includegraphics[scale=0.4]{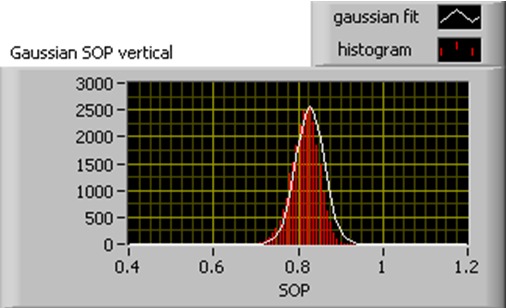} \label{gaussian_fits}
\caption{Guassian fits for individual peaks of above histogram of \ref{total_sop}} 
\label{histogram_with_fits}
\end{figure}

We then transmitted the random sequence of 0's and 1's for about 50000 data bits, SOP was computed for each bit using formula \ref{sop_apd}. A histogram of different SOP values were obtained and shown in figure \ref{SOP_histogram}. It shows two peaks, centered at $\pm 1$ respectively. It also shows that the distributions are not overlapping. In addition, these two peaks were fitted independently to Gaussian functions, in order to obtain the width of these distributions (figure \ref{histogram_with_fits}).

The width is a measure of polarization impurity,  since pure polarized light should give a very narrow distribution. The impurity is either due to actual degradation of polarization or could also be due to measurement errors and dark counts of the APD's. A very badly degraded polarization, or bad measurement would give rise to overlapping regions of the SOP peaks, indicating areas of erroneous bit assignment. 

\subsection*{Q factor and BER}
Even in the traditional On-Off keying, return-to-zero method, the photodiode measurement offers a similar distribution, giving some  nonzero value even when the signal is 0. This  can be fitted to a Gaussian function \cite{john}. The distribution would then consist of photodiode current when pulse is on (bit 1) and when pulse is off (bit zero). For such situation a Q factor is defined as \cite{vorontsov, john}
\begin{equation}
Q=\frac{I_1-I_0}{\sigma_1+\sigma_0},
\end{equation}
where $I_1$ is the center for Gaussian for `on' state and $I_0$ is the center of Gaussian for `off' state. $\sigma_1$ and $\sigma_0$ are the corresponding widths of the Gaussian. However, an equivalent Q factor can not be defined for our case. Since is Q factor is actually a measure of the imperfections, it can be defined as 

\begin{eqnarray}
Q_H&=&\frac{APD_H-APD_V}{\sigma_H+\sigma_V};~~~{\rm For~only~Horizontal~Transmission} \cr
\cr
\cr
Q_V&=&\frac{APD_V-APD_H}{\sigma_V+\sigma_H};~~~{\rm For~only~Vertical~Transmission}. 
\end{eqnarray}

The values $APD_V$ in first case and $APD_H$ in second case represent leakage and dark count values and corresponding $\sigma_V$ and $\sigma_H$ values represent the width of the histograms.  Hence the imperfections are accounted for in this measure. Although a perfectly symmetric system with respect to polarization should exhibit equal values for $Q_H$ and $Q_V$, most commercially available PBS show a little imperfection and hence slightly different values for $Q_{V,H}$. Overall Q factor can be taken as the average of these two. 

Again following the traditional approach as derived for On/Off keying, the theoretical bit error rate BER can be wrtiten as 

\begin{equation}
{\rm BER}=\frac{{\rm erfc}(Q\sqrt{2})}{2}.
\label{BER}
\end{equation}

But this will be the BER per vertical or horizontal polarization rather than the overall BER. With this theory, we do not have a calculation for the complete BER. 

In the next sections, we describe our results for simulated atmospheric conditions.

\subsection{Smoke}
Smoke was created in the chamber by burning household incense stick inside chamber. This created dense but lightweight smoke particle which hung inside chamber for sufficient time. Different amount of smoke was created by burning the stick for different amount of time, and the amount of smoke was quantified by the optical density, as given by equation (\ref{optical_density}). About 50,000 data bits were transmitted for each bunch and corresponding Q and BER for each polarization was computed. 

\begin{figure}[!h]
\includegraphics[scale=0.25]{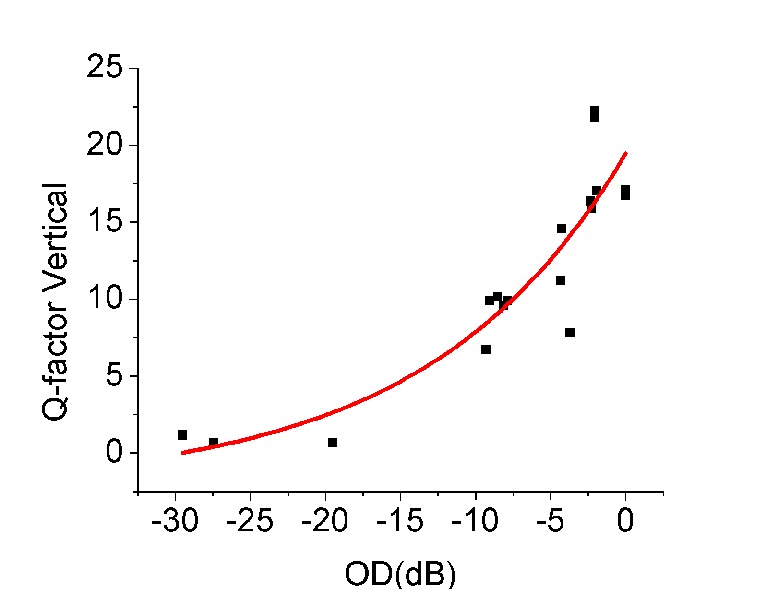}\hskip1.5cm \includegraphics[scale=0.25]{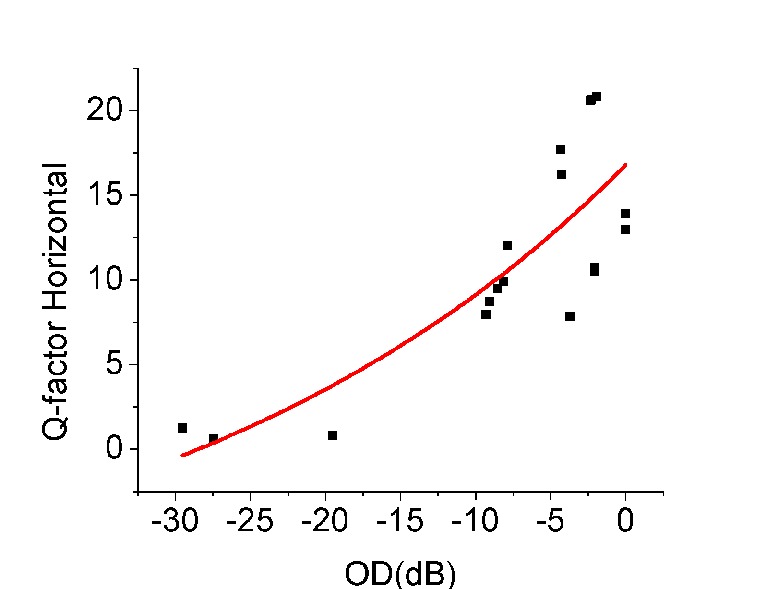}
\caption{Q factor for horizontal polarized (left) and vertical polarized (right) lights, in presence of smoke, as a function of Optical Density. Thick line is a curve fitting for exponential function.}
\label{qfactor_smoke}
\end{figure}

Figure (\ref{qfactor_smoke}) shows Q factor for horizontal and vertical polarized lights in presence of smoke. Optical density as defined by equation (\ref{optical_density}) is used for quantifying the amount of smoke in the chamber. It can be noticed that as Optical density increases from 0 to -30 dB, the quality factor Q reduces to almost zero. This is expected since the multiple scattering by the smoke particles degrades the polarization and errors increase. We fit an exponential curve to fit the data, indicated by thick line and shows a good agreement. 

\begin{figure}[!h]
\includegraphics[scale=0.25]{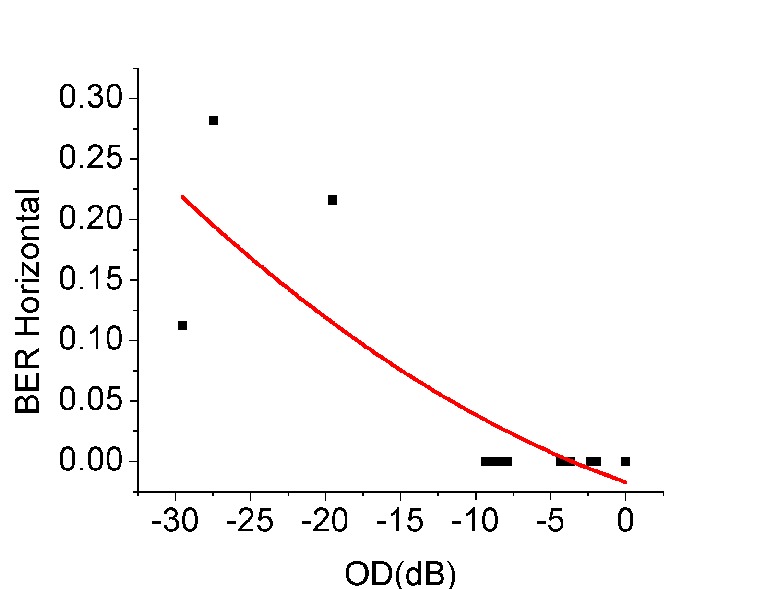}\hskip1.5cm \includegraphics[scale=0.25]{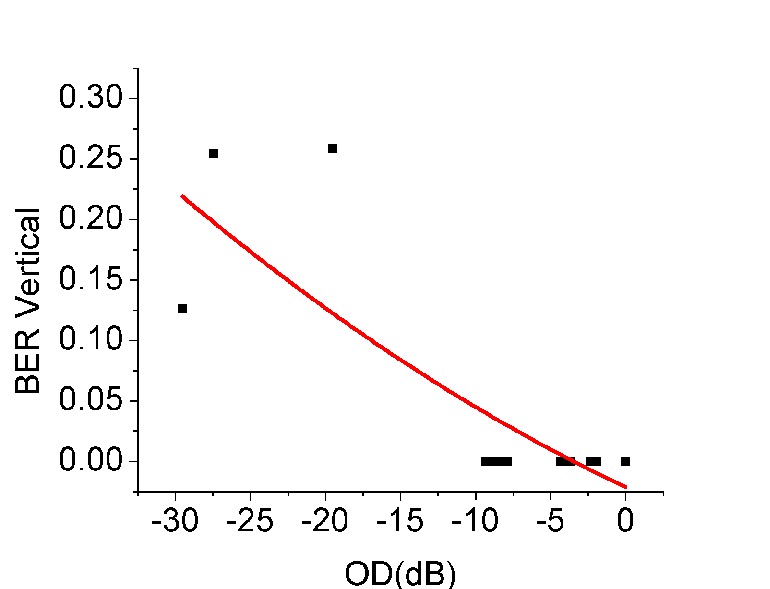}
\caption{BER as computed by equation \ref{BER} for  horizontal polarized (left) and vertical polarized (right) lights, in presence of smoke, as a function of Optical Density}
\label{smoke_ber}
\end{figure}

Figure (\ref{smoke_ber}) shows  corresponding bit error rate, as given by equation (\ref{BER}). The theoretical  BER is almost zero for small OD's but then increases rapidly for higher ODs. reaching almost equal to 25\% as optical density increases to -30 dB, which would be a significant amount of scrambling.    

\begin{figure}[!h]
\subfigure[]{\includegraphics[scale=0.5]{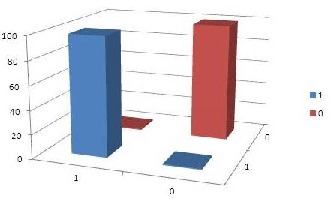}\label{correlation}}\hskip1.5cm\subfigure[]{\includegraphics[scale=0.22]{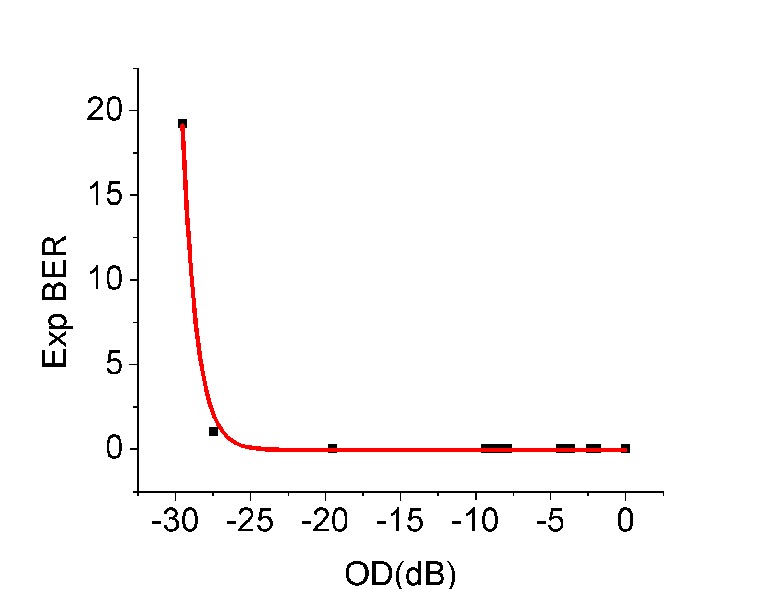} \label{smoke_ber_expt}}
\caption{(a) Correlation between transmitted bits, obtained from physically comparing each bit. The cross-correlation is almost zero and all bits are therefore transmitted with a very high fidelity. (b) Experimentally measured BER by comparing individual bits of transmitted and received data. Thick line represents a fit to exponential function  }
\label{sop_at_smoke}
\end{figure}

As earlier, SOP was computed for each bit. A histogram of the SOP distribution shows that its range is reduced from $\pm 1$ to a lesser range (figure \ref{sop_at_smoke}). However, the distributions are still not overlapping and hence it can be safely assigned to bit 0 when SOP is negative and to bit 1 when SOP is positive. 

After transmission, each bit on receiver was compared with the transmitted data and correlation  between them was computed. figure \ref{correlation} shows that despite high attenuation of light and a high degradation of the initial polarization, the actual bit error are almost zero and fidelity of transmitted data is very high, as can be seen by $1-1$ and $0-0$ correlations. 

But at higher concentrations of smoke, errors do develop, as shown by figure \ref{smoke_ber_expt}. It can be noted that actual BER is nearly zero for a very high amount of smoke, and suddenly increases only at a very high OD, like -30 dB. This shows that despite theoretically low values of Q and high values of BER, the actual BER is very less. It can be noted that theoretical BER is computed for On/Off keying while the experimental value is for PolSK, indicating that PolSK is much more robust compared to OOK. At very high OD, the error is more due to the very high attenuation wherein the APD's read zero counts rather than due to polarization scrambling. 

\section*{Fog}
Fog is an atmospheric condition made up of tiny water droplets, which also scatter light. Attenuation due to scattering, rather than absorption as in case of smoke is the key issue in case of fog. Multiple scattering by fog droplets also degrades polarization. Fog particles are larger than smoke particles and hence scattering cross section is higher. The polarization changes due to scattering is also different from that for smoke. 

We created fog in the chamber by sprinkling water over dry ice. As in case of smoke, a random sequence of 50,000 bits were transmitted using two VCSELs and the SOP for received data was computed. The Q factor and resulting BER were also obtained in similar fashion. 

\begin{figure}[!h]
\includegraphics[scale=0.2]{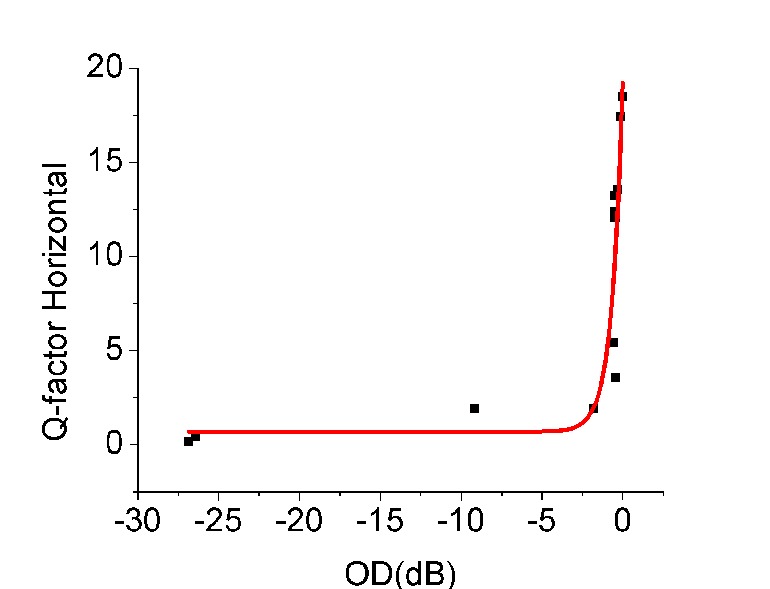}\hskip1.5cm \includegraphics[scale=0.2]{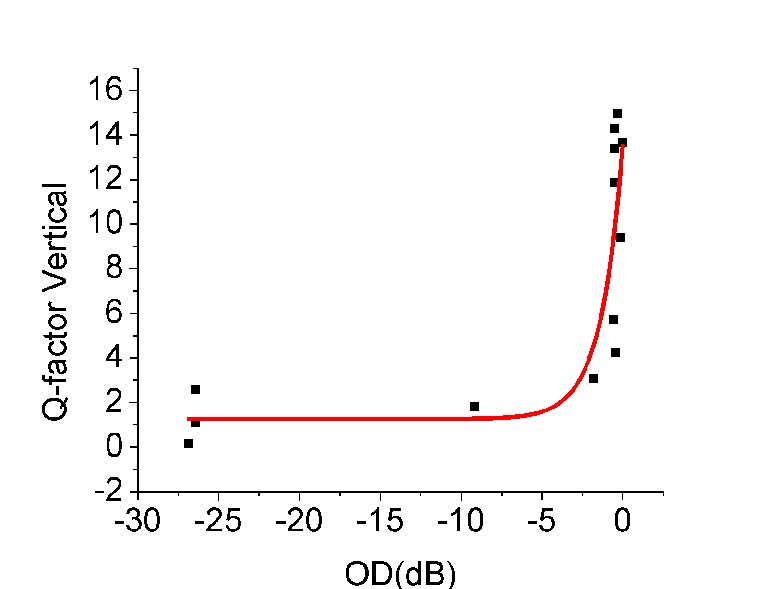}
\caption{Q factor for horizontal polarized (left) and vertical polarized (right) lights, in presence of fog, as a function of Optical Density}
\label{qfactor_fog}
\end{figure}

\begin{figure}[!h]
\includegraphics[scale=0.2]{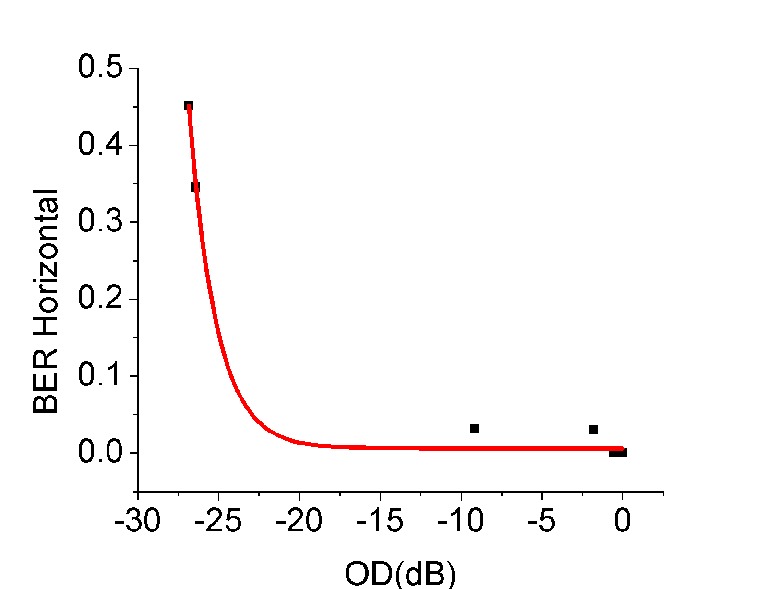}\hskip1.cm \includegraphics[scale=0.2]{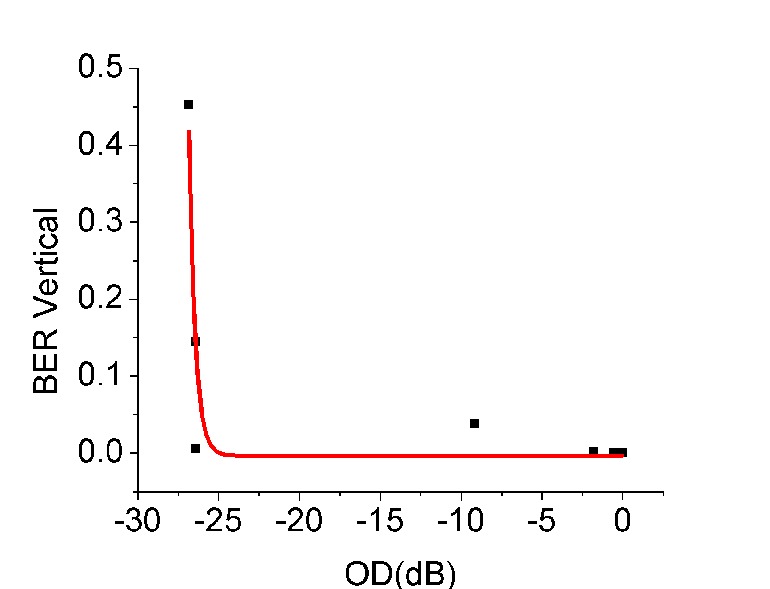}
\caption{BER as computed by equation (\ref{BER}) for  horizontal polarized (left) and vertical polarized (right) lights, in presence of fog, as a function of Optical Density. Solid line indicates a curve fit for exponential function.}
\label{fog_ber}
\end{figure}

Since the scattering characteristics and size distributions are different for fog and smoke, the theoretical Q and BER behaviour are also different. Q reduces at much faster rate and reaches zero even at OD of -10 dB. Theoretical BER, as computed by OOK formula reaches nearly 50\% for higher OD's, indicating a complete loss of information. 

As in case of smoke, we also calculated BER from comparing actual transmitted and received data. Although in case of smoke, the experimental BER for PolSK was much smaller than the theoretical BER for OOK, fog data on the other hand shows much closer comparison. This is because the scattering loss for polarization are different for smoke and fog particles. 

\begin{figure}[!h]
\includegraphics[scale=0.2]{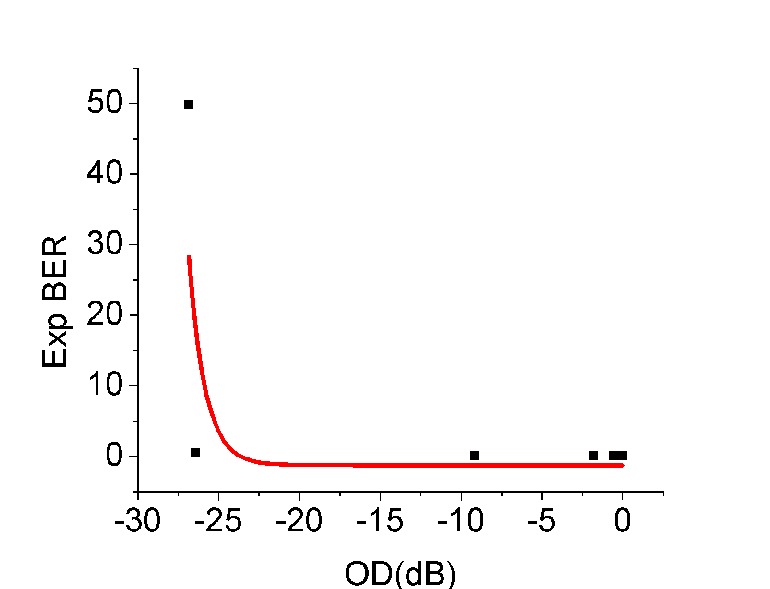}
\caption{Experimentally verified BER by comparing each bit. Solid line indicates a curve fit for exponential function.}
\label{fog_ber_expt}
\end{figure}

\section{Conclusions}
We have simulated atmospheric conditions inside laboratory to understand noise characteristics in Polarization encoding scheme. We have used vertical and horizontal polarizations to encode message bits. We have also used an Avalanche Photodiode based SPCM to increase sensitivity and work at regime of low incident intensities. Instead of a standard Stokes vector measurement, which requires six different measurements to analyse the state of polarization, we propose a more practical method of using only two measurements, thus reducing the number of detectors and therefore detector noise problems. However, the method proposed works only for horizontal and vertical polarizations and will not work for other modes of polarizations. 

We define a State-of-Polarization ${\cal S}$ based only two polarization states and  show that our method of assigning message bits 0 and 1 to negative and positive values of ${\cal S}$ respectively are more practical and show a very low bit error rates even in presence of thick smoke or fog. We compare the bit error for PolSK to analytical bit error rate of OOK scheme and show that PolSK is more robust and lower BER's than OOK, even when theoretical Quality factor is very close to zero. 

Although present data is for simulated conditions indoor, the  proposed system is easily extendable to outdoor situations. The proposed data analysis is easier requiring much lower resources such as detectors and electronics compared to other schemes. 

\section{Acknowledgement}
We thank Department of Information Technology, Govt. of India for financial assistance through their research grant. R.S. acknowledges the University Grants Commission, India for Rajiv Gandhi National Fellowship.

\end{document}